\documentclass[aps,preprint,groupedaddress]{revtex4-1}

\usepackage{amsmath}
\usepackage{amssymb}
\usepackage{graphicx}
\usepackage[version=3]{mhchem}

\begin{document}

\title{On the crucial features of a single-file transport model for ion channels}
\author{K.K.~Liang}
\email{kkliang@sinica.edu.tw}
\affiliation{Research Center for Applied Sciences, Academia Sinica, 128 Academia Road, Section 2, Nankang, Taipei 11529, Taiwan}
\affiliation{Department of Biochemical Science and Technology, National Taiwan University, No. 1, Sec. 4, Roosevelt Road, Taipei, 10617 Taiwan}

\begin{abstract}
It has long been accepted that the multiple-ion single-file transport model is 
appropriate for many kinds of ion channels.
However, most of the purely theoretical works in this field did not capture all of the
important features of the realistic systems.
Nowadays, large-scale atomic-level simulations are more feasible.
Discrepancy between theories, simulations and experiments are getting obvious,
enabling people to carefully examine the missing parts of the theoretical models and methods.
In this work, it is attempted to find out the essential features that 
such kind of models should possess, in order that the physical properties of an
ion channel be adequately reflected.
\end{abstract}
\maketitle

\section{Introduction}
The mechanism with which an ion channel conducts the ions through it had been the
aim of a tremendous amount of researches since the time of Hodgkin and Keynes%
\cite{hodgkin1955a}.
It will not be unfair 
to say that due to the importance of this problem,
the subject of \emph{single-file transport} made it to the center stage of theoretical biophysics.
Besides the purely theoretical interests of it, understanding of the transport mechanism is also
useful in many practical aspects.
The complicated current-voltage (I-V) relations of different ion channels,
the different manners in which they can be gated and regulated,
together with the selectivity of them, imply that they are like contemporary semiconductor
electronic devices that can be the logical operation components in complicated circuitries.
They are even more complicated than semiconductor devices because there are more than
two kinds of charge carriers in a biological ionic circuitry,
and each of them are regulated by many others in many ways.
Knowledge with which one can tune the electrophysiological characters of one of the devices
may enable one to fix malfunctioning biological circuitry.
Synthetic biologists may even design components with desired characters to implement
artificial organic machines\cite{l:inpress,acs:inpress}.
Structural biological studies provide quantitative information based on which
in depth studies can be performed.
Combinations of biophysical experiments and mutagenic manipulations
perturb the system in well-controlled manners to extend the dynamic range  of investigations.
It is hoped that theoretical studies can join the parade and provide solid understanding
and models with high predicting power.

The seminal works of Hille were perhaps the prototype of most, if not all, of the single-file
transport models of ion channels\cite{hille1975b,hille1978a}.
The channel is modeled by a chain of saturable binding sites.
It can be occupied by a number of ions.
The ions move in single file, that is, they do not penetrate through one another.
Each ion moves to one of its immediate neighboring vacant sites with absolute rates
determined through Eyring equation by the barrier heights between sites, 
the membrane potential, and the ion-ion repulsion.
Many works followed this idea and the early ones were summarized in
several good reviews\cite{levitt1986,nm5:1105,hille2001}.
Specifically for the stochastic models, there is an excellent recent review\cite{rmp85:135},
although the subject of ion channels is only a small portion of that review.
Recent works that are not categorized into the stochastic models
and are not yet reviewed include, but not exclusively, the works
from several groups that focused on continuum models\cite{%
bezrukov2005a,bauer2006,bezrukov2007a,ambjo2008a,ambjo2008b,prl100:038104,%
zilman2009}
and some others that transformed Hille's picture into exclusion-process models\cite{%
schumaker1990,chou1999a,chou1999b,grabe2006a}.
It is worthy of noting here that a major common technique in these theories is to use
steady-state fluxes as the quantities to compare with the observed ion currents.

However, some features of the above abstract models are not satisfactory.
Most of these works ignored Coulomb interaction between the permeating particles.
The particles are either non-interacting or the interaction is short-ranged.
Interestingly, this approximation rarely introduced troubles because in most of these works
only one particle was allowed to move in the channel pore each moment.
On the other hand, very strong Coulomb repulsion between ions had indeed been considered,
but that was done also by assuming that only one particle can be inside the channel pore
at a time\cite{prl100:038104}.

The interaction of the particles with the channel pore is an even more intriguing problem.
In the abstract models the potential surface seen by the permeating particles should,
of course, be artificially defined.
In order to mimic the effect of binding sites, it was invariably assumed that
local minima of the potential along the pore
that are able to tentatively trap the particles were present.
In exclusion-process models, these traps are further simplified into more abstract `cells'
that house the particles tentatively.
It was rarely reflected if the structure and therefore the potential
has anything to do with the presence (or absence) of the permeating particles.
The potential of mean force (PMF) experienced by the permeating particles, by definition, 
is averaged over the degrees of freedom of the remaining parts of the system,
including the protein, the membrane, the solvent, and other ions.
In principle it can be a static function of the space.
The problem, however, is whether 
the more relevant reaction path were chosen and how the total free energy of the system
was calculated.
This point will be further discussed later.


Thanks to the progress in structural biology studies%
\cite{jgp115:269,zhou2001b,nature417:515,nature417:523,nature423:33,science309:897,nature450:376,nature471:336},
nowadays atomistic simulations can be performed to study the transport processes with
increasingly realistic details included%
\cite{bpj78:2900,roux2001a,pnas100:8644,pnas101:117,bpc124:251,bpj90:3447,%
jacs130:3389,jmb389:637,psfb74:437,pnas106:16074,pnas107:5833,bpj101:2671,%
bc51:1559,jgp137:405,bpj103:2106,pcb8:e1002476,jgp141:619}.
Recently there was a review of the computational studies\cite{cr112:6250}.
Many other works mentioned in that review, especially those that were
not directly related to ion transport, were not cited in the above list.
Besides, several of the works mentioned above were newer than that review.
It seemed that atomistic simulations are now capable of replacing abstract theoretical models.
However, for many practical applications, abstract theoretical models are still more
convenient than simulations, when efficiency is considered.
Moreover, even though atomic-level details are included as much as possible,
the designs of these computer experiments were still not perfect 
in a few subtle yet crucial aspects.
Those are the focus of this article.

When the structures of KcsA at different potassium ion concentrations 
were solved, it was specially emphasized that\cite{zhou2001b} 
``in the selectivity filter a large number of negatively-charged carbonyl oxygen atoms 
point into the channel pore, 
making the pore very unlikely stable structure, unless several potassium ions are in the pore.''
Indeed, it is natural to suppose that in order to form the native structure those few potassium
ions observed should be in the channel at the right places at the beginning,
and the binding would be rather strong.
The entrance of a few ions into an empty channel pore
seems to happen routinely during reactivation after deactivation in
simulations on gating process\cite{science336:229}:
When the voltage-gated potassium ion channel was reactivated, 
water molecules and potassium ions flew into the channel 
to re-establish the conducting structure,
in early time in a molecular dynamics simulation, 
before permeation processes could have been observed.
Conversely, the channel stabilized the configuration of several positive ions,
otherwise the repulsion between ions would make it impossible for them to line up 
in single file to move across the membrane%
\cite{roux2001a}.
In other words, the structural features seem to suggest that when the channel is conducting, 
a number of ions have to be in the channel to stabilize the structure,
and the really meaningful potential surface for ion conduction
should be established inside the channel together with these first few ions.
The particles that enter the channel during conduction process will probably only perturb the
overall channel-plus-ions configuration relatively weakly to generate the ion current
through the so-called \emph{knock-on} type of transfer\cite{hodgkin1955a}.
If that were the case, question arises as what ``the reaction path for single file transport'' 
will look like.
That is equivalent to asking whether we are looking at one ion moving
between bulk solutions at both ends through the channel,
or we are looking at the transformation of the whole system 
with one more ion in the bulk solution on one side as the initial state, 
and the state with one more ion in the bulk on the other side as the final state.
In the latter case, obviously, 
the reaction path is generally different from the spatial coordinate along the channel pore.
Indeed, this is why nowadays simulation scientists emphasize more and more on
multiple-ion potential of mean force\cite{roux2001a,pnas100:8644,pnas106:16074,%
bpj101:2671,jmb401:831,jpcb114:13881}.


In this work, it is intended to set up a model in which ion-ion repulsion is included.
Besides, it is also intended that the role of the permeating ionic species in 
stabilizing the channel structure is explicitly considered.
Moreover, it is hoped that the model can be as simple as possible.
The purpose is to qualitatively demonstrate the impact of these effects on the
current-voltage-concentration relation of an ion channel%
\cite{pnas107:5833,bpj101:2671,jgp141:619}.
To that end, a chemical kinetics model is employed.
The channel is thought of as an enzyme which catalyzes the uptake of one ion on one side
of the membrane and the release of one ion to the other side of the membrane.
Therefore, the model involves the binding of ions into the channel as well as the
transport of the ions out of the channel.
The major features are as following.
First, the rate in which an ion exits the channel increases with the number of ions inside
the pore.
In other words, the exit of ions is a positively cooperative process.
Second, the binding constants of the first few ions into the channel are high,
but the binding constants decreases drastically when the number of ions further increases.

However, a simple model of ion transport is still rather difficult to handle,
not because the chemical kinetics model is difficult to set up, but because
approximations are difficult to apply to get results that are simple to manipulate.
Therefore, instead of directly studying the ion current of a channel system,
the unblocking kinetics of a blocked channel is studied\cite{jgp92:549,jgp92:569}.
In a previous work\cite{chang2009}, 
the kinetics of ion transport in the inner vestibule of inward-rectifying potassium ion channel 
Kir2.1 had been studied by blocking the selectivity filter part of the channels 
with barium ions and analyzing the exit rate of the barium ions. 
Compared to the apparent rate for a potassium ion to traverse the channel pore,
the rate in which the blocking barium be ``kicked out'' under the action of both
the membrane potential and the intra-channel potassium ions is much slower.
The slowing down of potassium ion motion permitted higher-quality experimental observations.
For setting up the theoretical model, there are also great advantages.
If the potassium ions will exit the channel more rapidly when the number of intra-channel ions
increases, so will the blocking barium ion exit more rapidly.
Indeed this was observed in the previous work.
More importantly, since the exit of the blocking barium ion is much slower than
the transport through an unblocked channel of a potassium ion,
the process of barium ion exit can be thought of as the rate-limiting step, and 
pre-equilibrium can be assumed for the processes of potassium ions binding into the channel.
If the potassium ion transport process were directly studied, 
probably the pre-equilibrium approximation cannot be applied.
If that is the case, instead of proposing several binding constants, 
a lot of rate constants have to be proposed, and the analysis will be distracted.
Nevertheless, two major difference between the unblocking and transport processes persist.
First, the unblocking process can only be better compared to the uni-directional transport.
The block-free transport is approximately uni-directional only when the
electrochemical gradient is steep.
Second, the blocking barium ion is positively charged and will influence the channel structure
non-trivially.
It is impossible to assess the influence of this fact on the soundness of the following discussions
in this work.
These problems can only be answered by further studies that take the suggestions given
in this work into account.

In the next Section, the details of the model is presented.
Most importantly, the relations between various binding constants and
rate constants are discussed carefully.
In Section~\ref{s:res}, simulations of the I-V curve as well as the concentration effect 
of the model system in different conditions are presented and the implications are discussed.
Then this work is concluded with some suggestions to both the simulation scientists
and theoreticians.

\section{Theory}\label{s:theory}

Consider a channel with $N$ possible ion-binding sites in it.
On one end of it, a barium ion can block the channel.
Upon referring to the above $N$ binding sites, 
the barium ion blocking site is not counted.
Potassium ions can enter from the other end of the channel pore.
Up to $N$ ions can be filled into the channel, if the physical condition permits.
In fact, it is not necessary to mention that there are $N$ binding sites.
Even if there is not any binding site or if there are a lot more binding sites inside,
the important feature is simply the maximum number of ions that can be in the
channel simultaneously.
In other words, for simplicity, the different binding states of a channel with $n$ ions inside
are considered indifferent in their effects on the barium ion exit rate.
In reality, the arrangement of the intra-channel potassium ions certainly have concrete
effects on the transport properties.
However, in the present abstract model, especially under pre-equilibrium approximation,
it is assumed that these details can be grossly lumped together.
Being thought about in this way, this model becomes exactly the same as the
sequential-binding model introduced by Weiss\cite{weiss1997}.
The sequential binding of potassium ions into the channel can be described by the
following chemical equations:
\begin{equation}\left\{
\begin{aligned}
\cee{K+(f) + Ba^{2+}(b) & <=>[K_1] K+(b) + Ba^{2+}(b)}\\
\cee{K+(f) + K+(b) + Ba^{2+}(b) & <=>[K_2] 2 K+(b) + Ba^{2+}(b)}\\
&\vdots \\
\cee{K+(f) + $\left(N-1\right)$ K+(b) + Ba^{2+}(b) & <=>[K_N] $N$ K+(b) + Ba^{2+}(b)}
\end{aligned}\right.
\label{e:rkbind}
\end{equation}
where K$^+$(f) means a free potassium ion in the bulk, while K$^+$(b) and Ba$^{2+}$(b)
means bound potassium and barium ions, respectively.
The equilibrium constants labeled above the arrows are binding constants of these reactions.
The subscript $j$ indicates that this is the process of the $j$-th potassium ion binding to
the channel.
Since in all cases one barium ion is in the channel, it can be neglected from the reaction quotient.
Similarly, in each of the reactions, the numbers of bound potassium ions in the initial and
final states of the reaction are fixed.
They can also be removed from the reaction quotient.
Eventually, the pre-equilibrium of this set of $N$ reactions is only determined by
the values of the binding constants and the concentration of the free potassium ions in the bulk.

The other set of reactions to be considered are the exit processes of
the barium ion under different conditions.
When there is not any potassium ions in the channel, according to the instantaneous
membrane potential, there is a probability that the barium ion will exit.
With the increase of the number of ions in the channel,
primarily due to the Coulomb repulsion, it can be expected that the exit rate of the
barium ion will increase.
These processes can be described by the following chemical equations:
\begin{equation}\left\{
\begin{aligned}
\cee{ Ba^{2+}(b) & ->[k_0] Ba^{2+}(f) }\\
\cee{ K+(b) + Ba^{2+}(b) & ->[k_1] K+(b) + Ba^{2+}(f) }\\
\cee{ 2 K+(b) + Ba^{2+}(b) & ->[k_2] 2 K+(b) + Ba^{2+}(f) }\\
& \vdots \\
\cee{ $N$ K+(b) + Ba^{2+}(b) & ->[k_N] N K+(b) + Ba^{2+}(f) }
\end{aligned}
\right.
\label{e:rbaexit}
\end{equation}
where Ba$^{2+}$(f) means free barium ion that just exits into 
the bulk solution on the opposite side of the channel.
The quantities $k_j$ over the arrow are the effective zeroth-order rate constants.
They are zeroth-order because each of them corresponds to a specific number of intra-channel
potassium ions, therefore this factor is already absorbed into the expression of the rate.
They are also only meaningful when the barium ion still blocks the channel.
Therefore effective zeroth-order rate constants are sufficient.
The subscript $j$ indicates the barium ion exit rate under the influence of $j$
intra-channel potassium ions.

Here it is emphasized again that since the barium ion exit processes, Reactions~\eqref{e:rbaexit},
are much slower than the potassium binding processes, 
pre-equilibrium approximation can be applied.
That is why in Reactions~\eqref{e:rkbind} only the binding constants are introduced,
while in Reactions~\eqref{e:rbaexit} the reaction rate constants are used to
describe the reactions.
Furthermore, in real experiments, after a preparatory step in which most of the
channels were found blocked with barium ion, the bulk solution was perfused so that
the bulk barium concentration was zero, before the unblocking experiment began.
Therefore, throughout the experiment, the bulk concentration of barium ions was so low that
the re-blocking of the channel was negligibly possible.
The rate constants of the reversed reactions do not have to be included in 
Reaction~\eqref{e:rbaexit}.

With the pre-equilibrium approximation, under a given condition,
the probability that there are $j$ ions in the channel can be easily derived\cite{weiss1997}:
\begin{equation}
p_j=\frac{\left(\prod_{m=1}^jK_m\right)\left[{\rm K}^+\right]^j}{
\sum_{n=0}^N\left(\prod_{m=1}^{n}K_m\right)\left[{\rm K}^+\right]^n}
\label{e:probj}
\end{equation}
where [K$^+$] is the bulk potassium ion concentration on the open-end side of the channel.
Besides the bulk concentration of potassium ion,
this probability also depends on $N$, but for brevity it is not explicitly labeled.
For example, if $N=3$ and $j=2$,
\begin{equation}
p_2=\frac{K_1K_2\left[{\rm K}^+\right]^2}{
1+K_1\left[{\rm K}^+\right]+K_1K_2\left[{\rm K}^+\right]^2+K_1K_2K_3\left[{\rm K}^+\right]^3}
\end{equation}
The expectation value of the barium exit rate under the given binding constants,
rate constants, and bulk potassium ion concentration, is
\begin{equation}
\bar{v}=\sum_{n=0}^Np_nk_n
\end{equation}
This is the major quantity of the theory.
It is directly proportional to the steady-state outward current.
Another quantity of interests is the mean number of potassium ions in the channel $\bar{n}$
given by
\begin{equation}
\bar{n}=\sum_{n=0}^Np_nn
\end{equation}
The explicit expressions of $\bar{v}$ and $\bar{n}$ are trivial but tedious, and are not shown here.
It has to be noted that since the barium exit rate is supposed to be proportional to the
potassium permeation rate and therefore it is used to represent the permeation rate in the
following, this model cannot be used to simulate negative ion current (or inward current).
Nevertheless, as long as the outward current is positive, the model may still work fine
even if the membrane potential is negative.
Therefore probably the reversal potential is still meaningful.

To perform simulations with this model, the relative values of the parameters
have to be determined.
Since the I-V curve is to be simulated, the dependence of the parameters on the membrane
potential $V$ also have to be determined.
Since the conventional definition of membrane potential is the intracellular potential
minus the extra-cellular potential, and since all of the ions explicitly considered are positive ions,
the open-end of the channel is considered to be intracellular and the barium-ion-blocked
end is extracellular.
In the following, the rules used in the present work are discussed.

When the membrane potential is zero, there is still a natural barium exit rate constant 
even without potassium ion in the channel.
This rate constant is labeled as $k_0^*$ (the star-sign indicates the field-free condition).
If one potassium ion entered the channel,
the barium ion exit rate constant will increase, and this increase is considered to be
related to the Coulomb interaction between the potassium ion and the barium ion.
With $j$ potassium ions in the channel, there are totally $j$ units of positive charge
repulsing the barium ion.
It is assumed that the activation energy of barium ion exit decreases linearly with
the number of potassium ions in the channel,
therefore the barium ion exit rate constant increases exponentially with the number
of potassium ions in the channel:
\begin{equation}
k_j^*=k_0^*\exp\left(\beta j\right)
\end{equation}
where $\beta$ is related to the sensitivity of the activation energy to the number of 
intra-channel potassium ions as well as temperature.

In the present model, it was proposed that the positive ions participate in stabilizing
the structure of the channel pore.
Therefore, the change in the number of ions in the channel affects the structure
or structural stability.
The barium exit rate will not only be influenced by the Coulomb repulsion.
Microscopic theories have to be used to find out the real form of its dependence on
the value $j$.
On the other hand, two properties assumed here are reasonable for the barium ion exit rate,
but are not really appropriate for describing the ion transport rate.
First, in the barium exit process, the extracellular concentrations of different ion species
are not very important.
Therefore the concept of reversal potential is not very relevant.
When the membrane potential is zero or even negative,
theoretically, there is always a finite barium exit rate.
However, for the ion transport process, there is always a thermodynamically well-defined
reversal potential, below which the macroscopic ion current turns zero and then negative.
In contrast, the barium exit rate is always positive.
Second, when the membrane potential is increased, the barium exit rate will
increase, and it diverges at some point.
This divergence is not strictly physical, but is understandable
because the barium ions will unblock the channel in almost no time.
However, typically, the ion transport rate will converge to a finite value\cite{bj41:119}.
Two reasons may be the most critical.
On one hand, due to the knock-on model of ion transport and the finite capacity of
the channel pore, at high voltage the channel is probably fully occupied and
is waiting for the bulk ions to knock on.
On the other hand, the friction due to ion-pore interaction and ion-ion repulsion
make the transport a drifting process.
There will probably be a terminal velocity.
Both ideas could lead to the same conclusion that the rate in which the intra-cellular
ions hit the entrance of the channel determines the maximal ion transfer rate.
Therefore it is reasonable that the observed asymptotic value of the I-V curve is
closely related to the diffusion rate of the permeating ion.
In this work, the discussions are focused on the change in the magnitude of the I-V curve.
Therefore, the discrepancy between this model and the experiments 
at the negative- and positive-voltage extremes will not be seriously treated.

In comparison, a more complicated dependence of the field-free binding constants
on the potassium ion number is proposed.
As mentioned in the Introduction, it is desired that the first few potassium ions bind with
higher affinity, while the binding constant decreases drastically for excess potassium ions.
In the following, the number of preferentially bound potassium ions is set to two.
To realize such a model, a Fermi-distribution-like dependence of $K_j^*$ on $j$
with the ``Fermi level'' at $j=\mu$ is proposed:
\begin{equation}
K_j^*=B/\left(1+e^{a\left(j-\mu\right)}\right)\label{e:Kjstar}
\end{equation}
where $a$ and $B$ are constants.
In the next Section, $B$ will be represented by another parameter $b$ such that $b=\ln B$
just for numerical convenience.
Notice that when $B$ is changed, 
all of the binding constants are changed by the same multiplicative factor.
The ratios between the magnitudes of the binding constants do not change with $B$.
This is certainly not the perfect model.
Nevertheless, depending on the emphasis of the study, it may be not a poor model, either.
Although the ratios of the form $K_i/K_j$ is independent of $B$,
when $j\lesssim\mu$,
the terms of the form $\prod_{m=1}^jK_m$ in Eq.~\eqref{e:probj} are proportional to $B^j$.
These quantities are of interests because $K_j=\exp\left(-\Delta G_{j,j-1}/RT\right)$ and 
$\Delta G_{j,j-1}=G\left(j\right)-G\left(j-1\right)$ where $G\left(j\right)$ is supposed to be
the absolute free energy of the system with $j$\/~potassium ions inside the channel.
Therefore $\prod_{m=1}^nK_m=\exp\left(-\left[G\left(j\right)-G\left(0\right)\right]/RT\right)$.
Defining $\Delta G_j=G\left(j\right)-G\left(0\right)$ then
$\prod_{m=1}^nK_m=\exp\left(-\Delta G_j/RT\right)$.
When $j\lesssim\mu$, the denomiator part of Eq.\eqref{e:Kjstar} is around 1, and
$\prod_{m=1}^jK_m$ is dominated by $B^j$.
Therefore the relative probability of having larger number of intra-channel ions 
increases significantly with $B$, until there are enough positive ions to stabilize the channel.
The free energy of the channel, compared with its empty state, decreases with
increasing number of ions inside.
For $j>\mu$, the denominator part of Eq.\eqref{e:Kjstar} increases rapidly with $j$,
and the product turns decreasing with $j$.
That is, the free energy of the system starts to increase with more ions inside.
These are indeed the qualitative features desired by the present model,
which are not trivial and difficult to embody quantitatively.
The good point is that only a very limited number of arbitrary parameters are introduced.
The parameter $a$ also controls how abruptly the binding affinity decreases 
when the intra-channel potassium ion number increases beyond $\mu$.
Throughout the following text, its value will be fixed at $a=2.5$, while $\mu=3$.

\begin{figure}[ht]
\begin{center}\includegraphics[width=6cm]{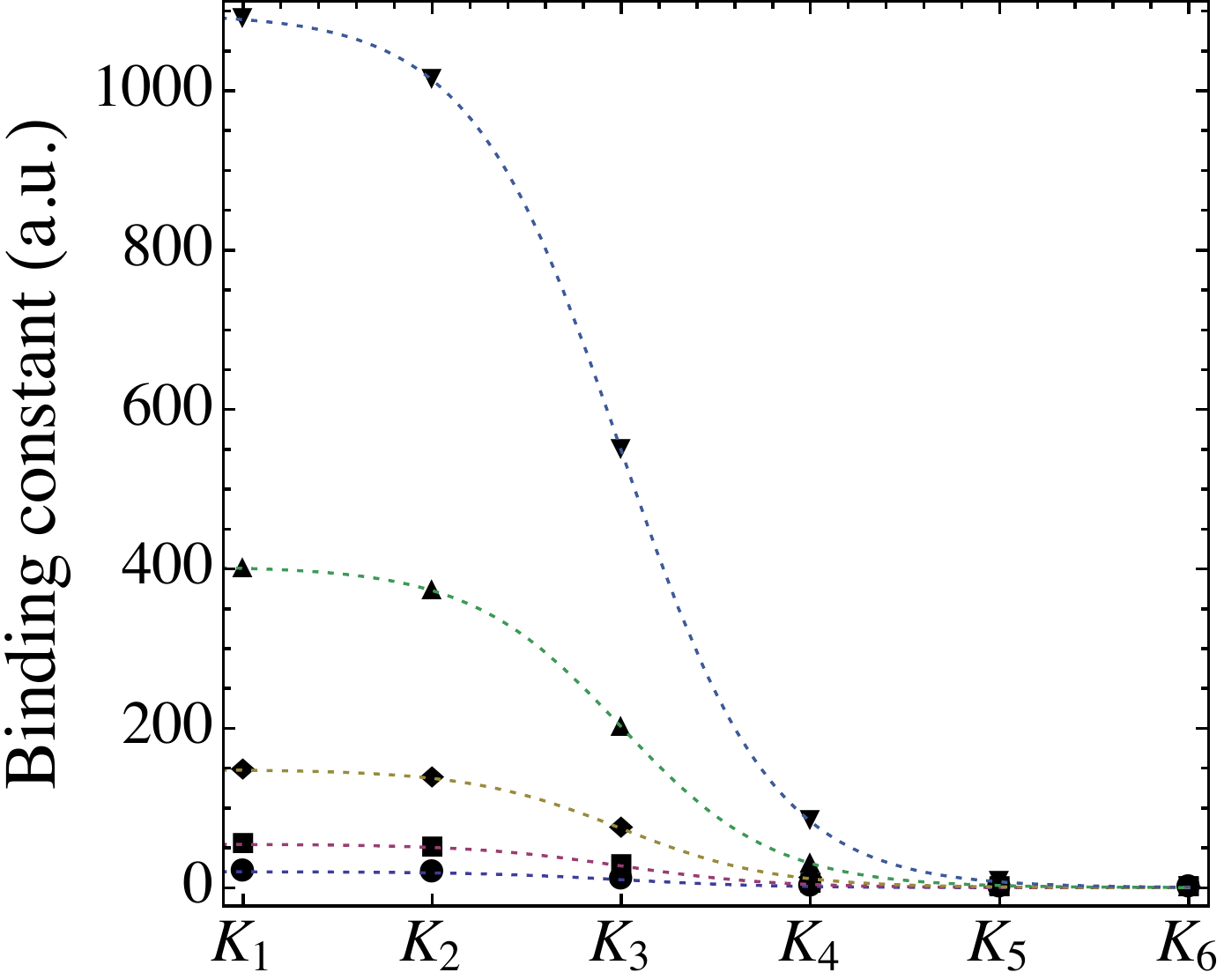}\end{center}
\caption{The values of $K_j$ for $N=6$, $\mu=3$ and $a=2.5$ are presented.
The filled circles indicates the $K_j$ values, while the dotted lines shows the trend.
The data shown, from the lowest to the highest curve, correspond to
$b+\delta V=3$, 4, 5, 6 and 7, respectively.
\label{f:fig01}}
\end{figure}

When the membrane potential $V$\/ is finite, the parameters have to be further adjusted.
The simplest exponential-form I-V curve commonly saw in microelectronics textbook is used.
Again, this is the proper model for barium exit and qualitatively related to the ion current.
By representing the sensitivity of the exit rate constants to the membrane potential and 
temperature with a factor $\kappa$, the full form of the barium ion exit rate constants $k_j$
in this model is
\begin{equation}k_j=k_0^*\exp\left(\kappa V+\beta j\right)
\label{e:baexitk}\end{equation}
The ion binding reactions are considered as barrier-crossing processes of charged particles.
The dependence of the activation energies on membrane potential is assumed to be linear,
from the most straight-forward and lowest-order energetic consideration.
Therefore, under the action of membrane potential $V$, the potassium ion binding constants are
\begin{equation}
K_j=e^{b+\delta V}/\left(1+e^{a\left(j-3\right)}\right)
\label{e:eqkdef}
\end{equation}
where $\delta$ represents the degree of sensitivity of the binding process to the
membrane potential.
Again, notice that when the membrane potential is raised,
all binding constant $K_j$ increases with the same proportion.
As discussed earlier, in the present model as long as the ion can enter the channel
it can be considered as bound.
Therefore in any case the increased membrane potential should shift
the binding equilibrium to the forward direction.
Thus this property is also qualitatively correct.
A plot of $K_j$ for the $N=6$ case is shown in Fig.~\ref{f:fig01} for reference.
In the next Section, both $N=6$ and $N=4$ cases will be simulated and discussed.
If all other parameters are kept the same, the first four binding constants $K_1$ through
$K_4$ will be the same in both cases.

In the following Section, selected results of numerical calculations based on this model
are presented.
Since $\bar{n}$ is obviously dimensionless, and $\bar{v}/k_0^*$ is also dimensionless,
actually the time scale, voltage, current, and ion concentration are all relative quantities.
Those results will only be discussed qualitatively.

\section{Results and Discussions}\label{s:res}

\begin{figure}[ht]\begin{center}
\begin{tabular}{cc}
(a) & (b) \\
\includegraphics[width=6cm]{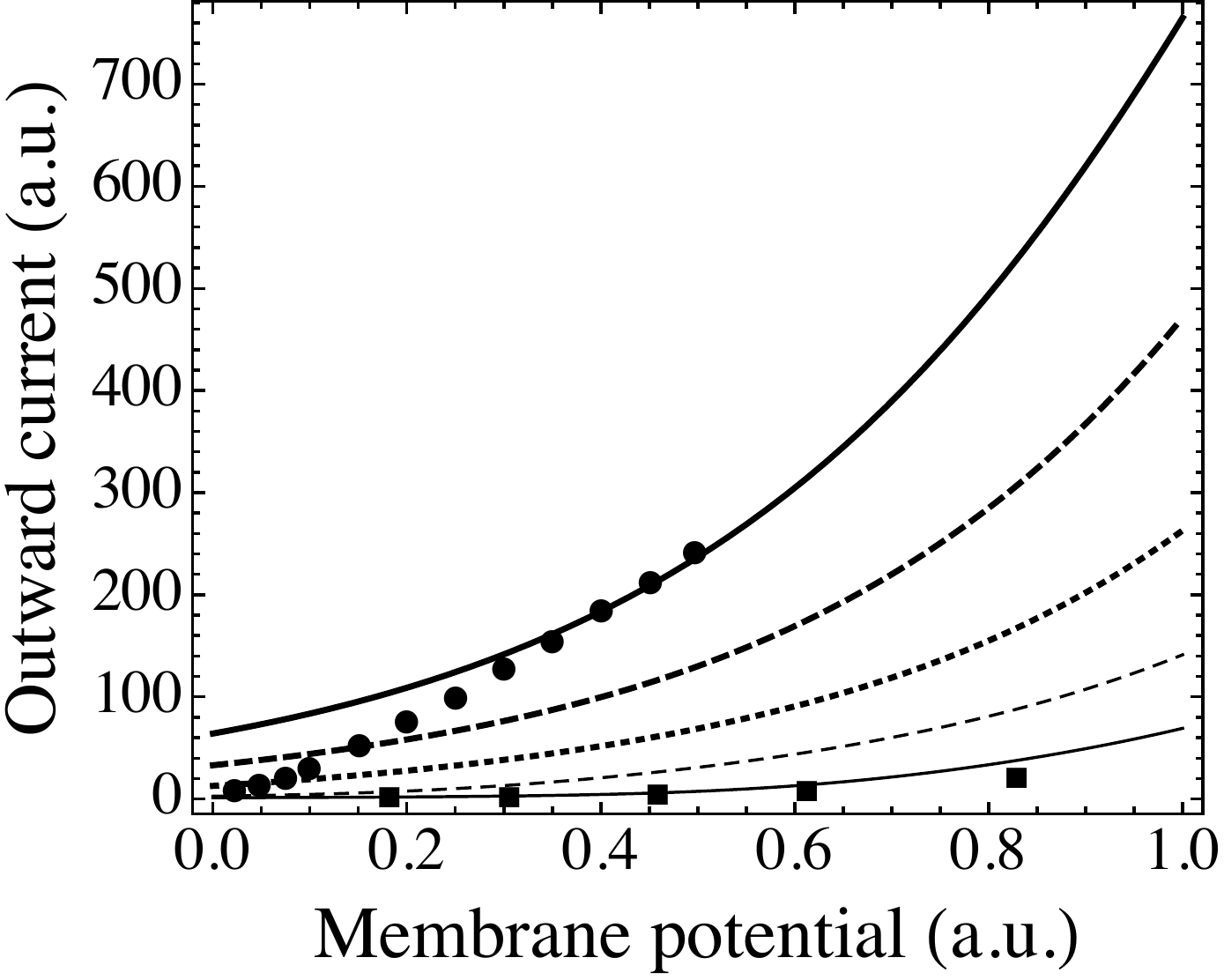} &
\includegraphics[width=6cm]{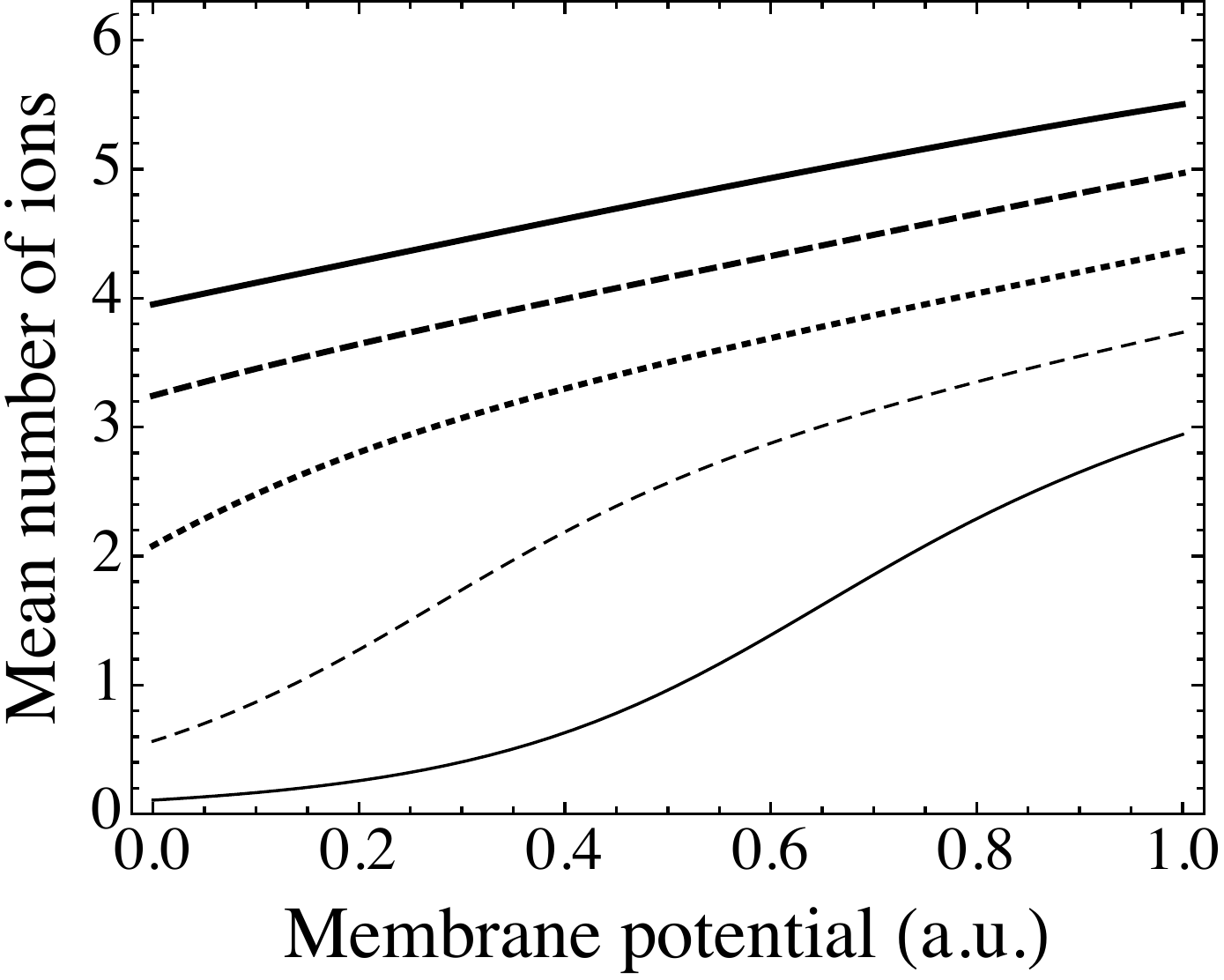}
\end{tabular}
\caption{The properties of a system of channels with six binding sites inside.
The I-V curves of the system, (a), and the mean number of potassium ions 
in the channel, (b), are plotted as functions of membrane potential.
The unit of the membrane potential is arbitrary, but the parameters were chosen
so that it seems the unit of the potential is Volt.
The five curves in each panel correspond to $b=0$ (thin line),
1.5 (thin dashed line), 3.0 (dotted line), 4.5 (thick dashed line) and 6.0 (thick line), respectively. 
In Panel (a), also plotted are the I-V relations measured experimentally (filled circles)%
\cite{bj41:119}\ %
and simulated by molecular dynamics simulations (filled squares)\cite{jgp141:619}.
\label{f:fig02}}
\end{center}\end{figure}

First, consider a channel with six potassium binding sites.
In the first set of results, the following values of the parameters are used.
The sensitivity factors of the potassium ion binding constants and 
the barium ion exit rate constants on membrane potential are
$\delta=4$ and $\kappa=1$, respectively.
The bulk potassium ion concentration is $\rm\left[K^+\right]=0.1$.
The sensitivity factor of the barium exit rate constant on
intra-channel number of potassium ions is $\beta=1$.
The cases corresponding to five different values of $b=\ln B$
from 0 to 6 in steps of 1.5 are presented in Fig.~\ref{f:fig02}.
In either Panels, the cases of $b=0, 1.5, 3, 4.5$ and 6 are plotted in
thin solid line, thin dashed line, dotted line, dashed line and solid line, respectively.
In Panel~(a), the I-V curves for positive membrane potential are presented,
where the voltage range is from 0 to 1.
The $y$-axis is the outward current, but actually what is plotted is the scaled
expected barium exit rate $\bar{v}$.
Also plotted are the data obtained from the literature.
The filled circles are the experimentally (single-channel recording) measured
I-V relation of gramicidin A channel by Anderson\cite{bj41:119}.
The filled squares are the simulated ion current by Jensen \emph{et al.}\cite{jgp141:619}.
The comparison between the present model with these data will be discussed later.
All of the theoretical curves look like exponential curves, but actually they are not.
The important feature is that, with increasing $b$ or $B$,
the whole I-V curve shifts in the direction of higher current.
To explain the trend of the I-V curves,
in Panel~(b) the mean number of potassium ions in the channel, $\bar{n}$,
is plotted with respect to membrane potential.
For zero or small values of $b$,
it is seen that at low membrane potential there is hardly any potassium ion
inside the channel.
However, $\bar{n}$ increases more rapidly after some turning point.
When $V$ is increased to about 1.0, the channels are averagely half-filled.
Therefore, the corresponding I-V curves have much smaller slopes at lower voltage.
For higher $b$ value, of course the number of ions in the channel is much larger
even at very small membrane potential, because the first few binding constants
are significantly higher.
However, since $K_5$ and $K_6$ are still very small, it is very difficult to saturate the channel,
and $\bar{n}$ only increases slowly.
But this is already sufficient to explain why the magnitude of the current is higher
with higher $b$ value.
Besides, if the current axis is logarithmic instead, the I-V curves have, respectively,
very similar shapes as their corresponding $\bar{n}$-V curves.
In other words, the upper-most curve $\left(b=6\mbox{ or }B=400\right)$ 
is close to exponential,
but the lowest curve $\left(b=0\mbox{ or }B=1\right)$ is highly non-exponential.

\begin{figure}[ht]\begin{center}
\begin{tabular}{cc}
(a) & (b) \\
\includegraphics[width=6cm]{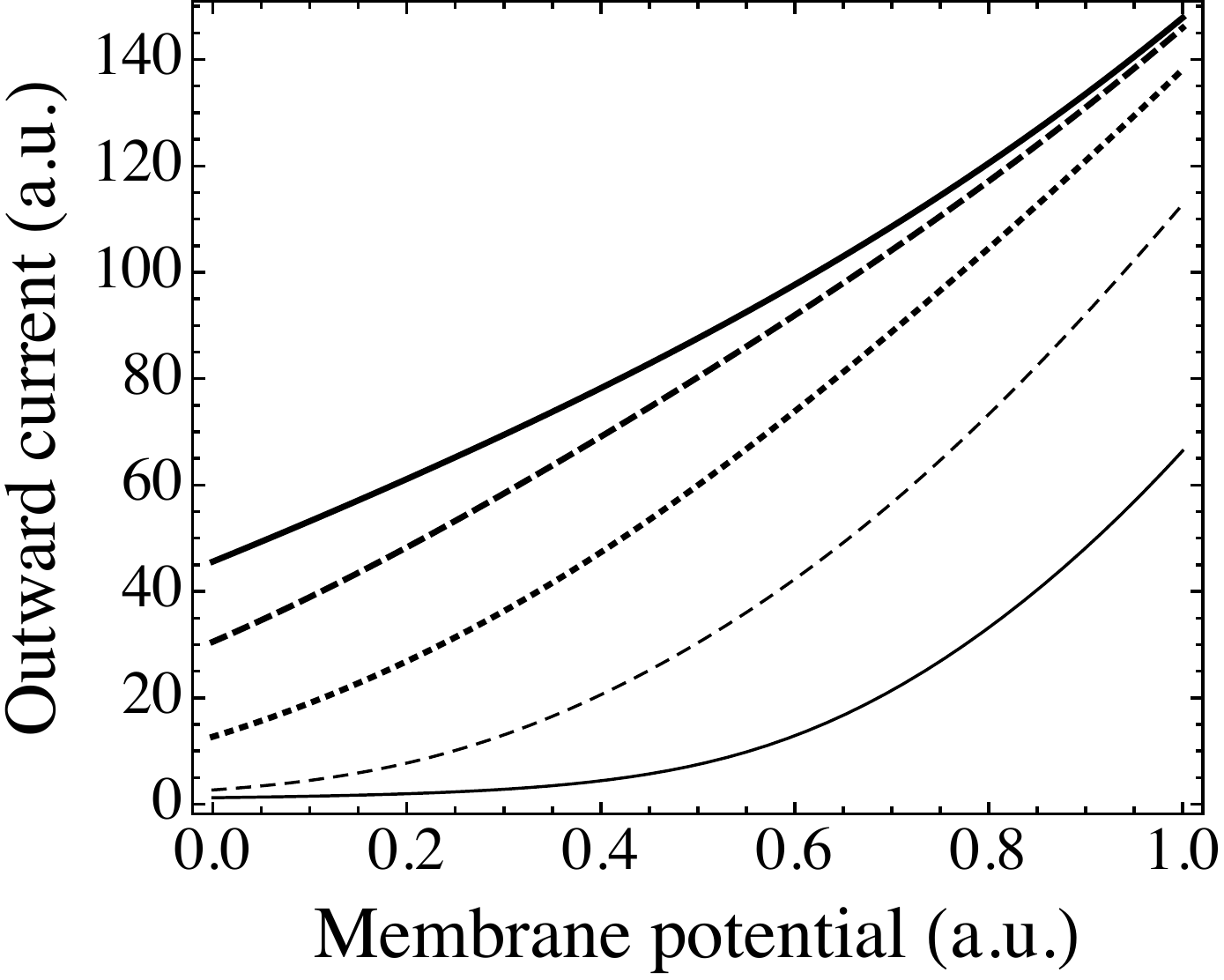} &
\includegraphics[width=6cm]{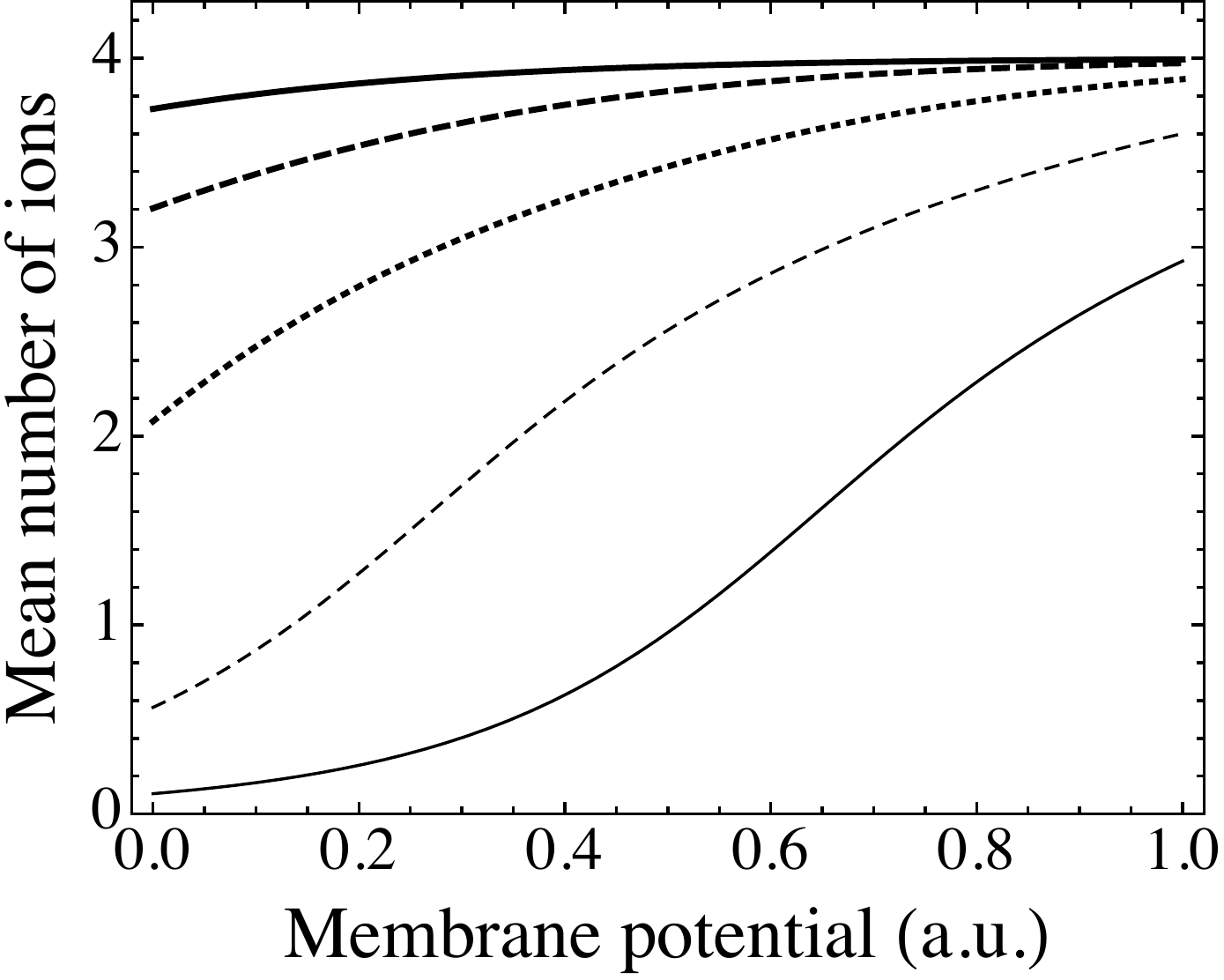}
\end{tabular}
\caption{The properties of a system of channels with four binding sites inside.
The I-V curves of the system, (a), and the mean number of potassium ions 
in the channel, (b), are plotted as functions of membrane potential.
The five curves in each panel correspond to $b=0$ (thin line),
1.5 (thin dashed line), 3.0 (dotted line), 4.5 (thick dashed line) and 6.0 (thick line), respectively. 
\label{f:fig03}}
\end{center}\end{figure}

The non-exponential features of the I-V curve is better seen if the case of $N=4$
is shown instead.
With all other parameters remaining the same, including $K_1$ through $K_4$,
the I-V curves and $\bar{n}$-V curves under the same conditions are shown in
Fig.~\ref{f:fig03}.
According to Fig.~\ref{f:fig01}, $K_3$ and $K_4$ are much larger than $K_5$ and $K_6$
when all of the parameters except $N$ are the same.
Therefore in the $N=4$ case, the channel is much more easily saturated.
Even when $b$ is large, unlike the situation in the $N=6$ system,
the $\bar{n}$-V curve is quite nonlinear.
Compared with Fig.~\ref{f:fig02}(b), it is obvious that this nonlinearity is mainly due to
the saturation of the channel.

Under high membrane potential, the I-V curves of the present model all become exponential
because the channel pore is nearly saturated with the permeating ions so that
$\bar{n}$ is nearly constant, but the barium exit rate $k$ increases exponentially with the
membrane potential so that $\bar{v}$ also increases exponentially.
As the discussion about Eq.~\eqref{e:baexitk} in the previous Section explained, in reality, 
the current will typically saturate at a value believed to correspond to the diffusion-limited
ion transport rate\cite{bj41:119}.

Next, the comparison with the studies on real ion channels is discussed.
In Fig.~\ref{f:fig02}a, the MD-simulated I-V curve\cite{jgp141:619}
and the experimentally measured results\cite{bj41:119} were shown.
The relative scale of these two sets of values were fixed, exactly as shown in
Jensen's comparison\cite{jgp141:619}.
However, to show them side-by-side with the present theory,
they were all multiplied by 1.2 times.
This made the simulation results comparable with the $b=0$ case in this theory.
In that case, it appears that the magnitude of the experimental result is
comparable with the $b=6$ curve of this theory.
The shapes of the curves are, however, not closed to each other.
Especially, in the present model, the $y$-intersect of the I-V curves depends on
the binding constants quite strongly.
In the experiments done by Andersen\cite{bj41:119}, all of the experiments
were done on non-blocked gramicidin A channels with equal salt concentrations
on both sides of the membrane (0.1~M) so that the reversal potential is always 0.
This difference between the present model with the experiments (and simulations)
has also been discussed in the previous Section.

Nevertheless, the discussion about the change in magnitudes of the currents is
still meaningful.
According to the design of this model and the discussion by Jensen,
it seems that this similarity in the relative change in magnitude is not accidental.
In the molecular dynamics simulation, it was proposed that the low-voltage I-V characteristics
is due to the ``infrequent ion recruitment into the pore lumen'' 
in the case of gramicidin A channel
and that the ``formation of the knock-on intermediate occurred too infrequently''
in the case of K$_{\rm v}$1.2/2.1\cite{jgp141:619}.
In the present simple model, the flat low-voltage part is 
simply due to the small number of potassium ions in the channel.
However, since it is assumed that two ions are much more preferentially bound,
the binding of exactly two ions in the channel may be the analogy of the formation of
the knock-on intermediate in K$_{\rm v}$1.2/2.1.
In the structures used for gramicidin A channel simulations (PDB 1JNO and 1MAG),
there is not any ion shown in the structure, but both structures were obtained from
crystals of gramicidin A channel complexed with potassium thiocyanate.
In the structures of Kv1.2/2.1 (2R9R and 2A79) or KcsA (3F7V and 1K4C),
there are potassium ions at specific positions in the structures.
It is reasonable to assume that during the experiments the structures of
these channels were stabilized by including permeating ions in them.
However, in the relaxed structures in the MD simulations,
it seemed that the structures were not stabilized by including these ions.
Some artifacts may be there, making it much more difficult for the ions to enter
the simulated channels.

\begin{figure}[ht]\begin{center}
\begin{tabular}{cc}
(a) & (b) \\
\includegraphics[width=6cm]{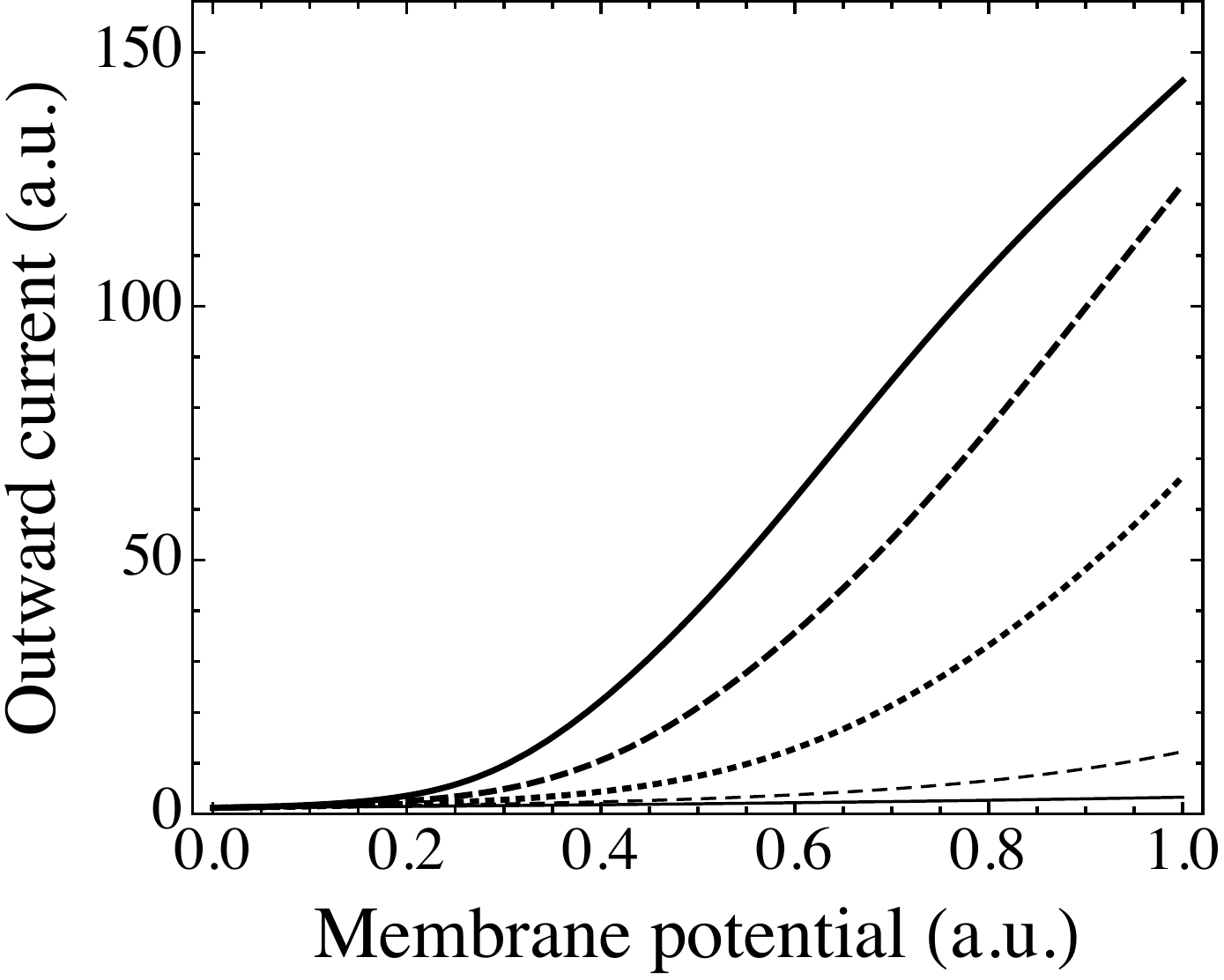} &
\includegraphics[width=6cm]{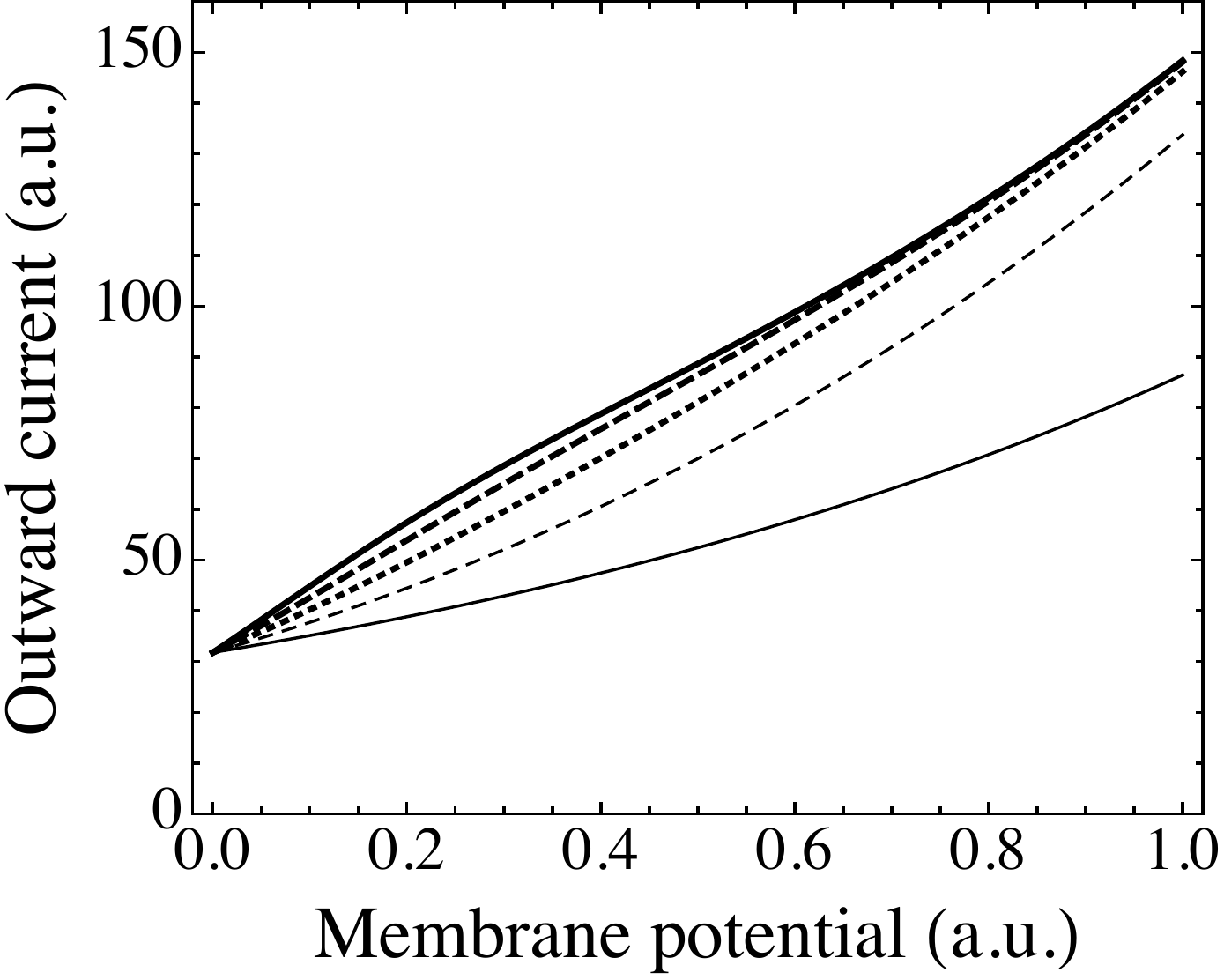}
\end{tabular}
\caption{The I-V curve of the model system under different conditions.
In Panel (a), the parameter $b$ is 0, or $B=1$, in Eq.~{\protect\eqref{e:eqkdef}«}.
In Panel (b), $b=\ln 100$ and $B=100$.
In each Panel, five different curves correspond to different values of the sensitivity, $\delta$, 
of the potassium ion binding constant to membrane potential: the thin line for $\delta=0$,
thin dashed line for $\delta=2$, thick dotted line for $\delta=4$,
thick dashed line for $\delta=6$, and the thick line for $\delta=8$.
\label{f:fig04}}
\end{center}\end{figure}

In Equation~\eqref{e:eqkdef}, the two factors $b$ and $\delta V$ appear together in the
same exponent.
It is necessary to further emphasize the different influence of these two terms.
Obviously, at low-voltage, the I-V characteristics is mainly determined by $b$,
among other constant factors.
If $b$ is relatively small, the high-voltage properties of the system is dominated by
the membrane potential.
To further distinguish the two effects, in Fig.~\ref{f:fig04}, the I-V curves at two different
$b$ values are presented.
In Panel~(a), $b=0$ and $B=1$. The five I-V curves correspond to different values of
the sensitivity $\delta$ of the binding constant to voltage: $\delta=0$ for the thin line,
2 for the thin dashed line, 4 for the thick dotted line, 6 for the thick dashed line, and
8 for the thick line.
In Panel~(b), $b=2$ and $B=100$.
The five I-V curves correspond in exactly the same way to the sensitivity factor $\delta$.
In both Panels, the maximum current values are about the same,
making the comparison easier.
When the value of $b$ is large, the effect of the membrane-potential-sensitivity is less obvious.
A drastic difference in the two cases is that, when $\delta$ and $b$ are both small,
the current is very low even at relatively high voltage;
but if $b$ is not so small, even if $\delta$ is zero, the I-V curve is obviously concave upward.
This is because when $b$ and $\delta$ are both small, even at relatively high voltage,
the mean number of potassium ions in the channel is close to zero.
Therefore, the majority part of the barium ion exit rate, that is, the so-called current that
is calculated in this model, is from $k_0$, which is pretty small.
Of course, if the value of the parameter $\kappa$ is increased, these few I-V curves
will not look so flat, but the fact that these curves are flat is exactly the point of this
model-system study.

By comparing with the experimental and simulation results presented by Jensen\cite{jgp141:619},
it appears that the experimental results are closer to the high-$b$ cases in this model,
while the simulation results are closer to the low-$b$ cases.
What can be the implications?
Since the simulated currents are too weak compared with experiments,
Jensen {\itshape et al.}\ tried to modify the ion-pore interaction to see if there can be
more ions in the channel so that the current can also increase.
This does not work for K$_{\rm v}$1.2/2.1.
For gramicidin A, there is increase in both intra-channel ion number and current
when the ion-pore interaction is increased, but still way too low.
Beyond these descriptions, there were too little information provided.
However, compared with the present model, it seems that the major difference between the
low-current and high-current cases are not solely in the attraction interaction between
the ions and the channel pore.
Indeed, increasing the attraction will lower the energy of the ions in the pore.
But according to the present model, the most significant effect would be due to
the better stability of the channel structure, instead of just from being able to
attract the ions closer to the channel.
If the channel structure is `relaxed' without putting a few positive ions inside,
extra forces must have been applied somewhere to make the structure more
stable than it should be.
And, since by doing that the channel is already `over' stabilized,
the entrance of more than one positive ions in the channel will most likely be
energetically unfavored.
In other words, if the simulation results can really be used to calculate whether the whole system
can go through a structure reorganization so that the free energy change for binding the
first few ions is indeed much more significant, that should help to greatly increase the
current level predicted by the simulation.
By reading their article, it is not explicitly found out whether those authors
included the potassium ions in the relaxation stage or not.
By studying their procedures of obtaining the one-dimensional potential of mean force
and the simulated I-V curves, it seemed that the initial structure of the channel was
devoid of potassium ions, but still this is not sure.
Even if potassium ions were included, it was very difficult to judge whether the
structure was fully relaxed or not.
The practical limitation is that in order to compare the stabilities of the channel structures
with different numbers of ions in them, very long simulations have to be carried out.
It takes strong motivation for the simulation scientists to devote computational resources
on such works.
Hopefully the discussions here provide some encouragement.



\begin{figure}[ht]\begin{center}
\includegraphics[width=6cm]{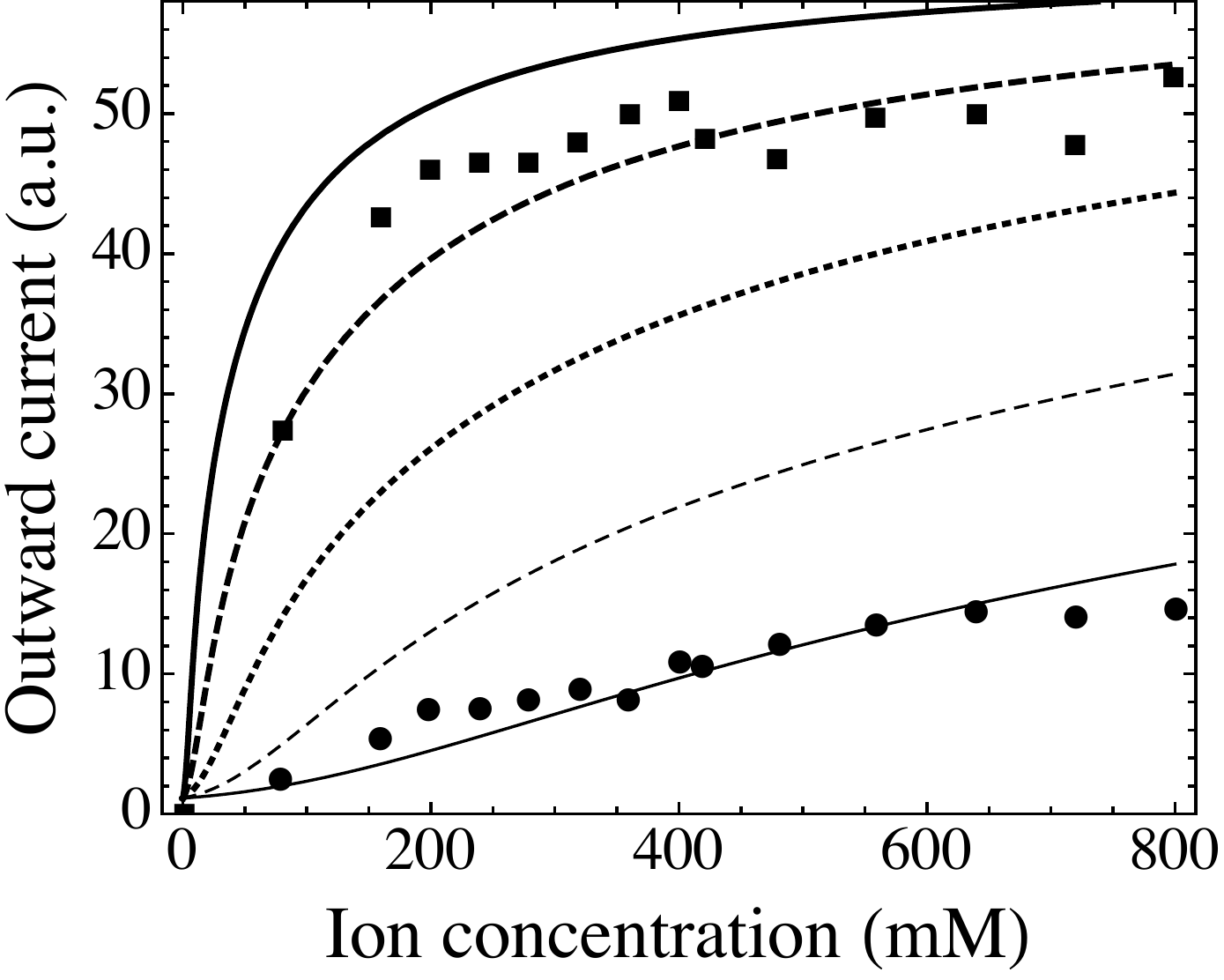}
\caption{The current-concentration relation of the present model is compared with
that obtained through Brownian dynamics and molecular dynamics simulations\cite{bpj101:2671}.
The filled circles are the simulated current-concentration relation of wild-type Kv1.2 channel.
The filled squares are the simulated current-concentration relation of mutated Kv1.2 channel
in which two of the proline residues were replaced by aspartate residues.
The current values of all of these data were multiplied by 6 so that they can be
in the same scale as the currents calculated from the present model.
The five lines are the current-concentration relation calculated from the present model.
The number of binding sites is assumed to be 4.
Other parameters are $a=2.5$, $\mu=3$, $\delta=10$, $\kappa=1$, $\beta=1$,
and membrane voltage $V=0.12$.
The five curves, from the bottom to the top, correspond to $b=0$, 1, 2, 3 and 4,
respectively.
\label{f:fig05}}
\end{center}\end{figure}

Besides the I-V relation,
it is also worthy of examining whether the dependence of ion current on the bulk concentration
of potassium ions shows reasonable trend.
For this purpose, comparisons were made with the Brownian dynamics simulations done by 
Gordon and Chung\cite{bpj101:2671} on Kv1.2.
It was argued in their work that the pore region of Kv1.2 is much more hydrophobic and
less charged compared with KcsA.
The low current level in the simulations on wild-type Kv1.2 was attributed to the
higher hydrophobicity and neutrality of the channel pore.
To examine whether the conjecture is reasonable, they mutated two proline residues into
aspartate residues to increase the Coulomb interaction between the potassium ions and the
channel pore.
From their figures, the current-concentration data points were extracted.
Since the current level is arbitrary in the present model,
the ion-current values taken from their data were scaled arbitrarily 
so that they can be directly compared with the results of the present model.
Moreover, since they explicitly used pico-ampere as the unit of the ion current,
it is assumed that the relative scale between their data is absolute.
The factors used for scaling the two set of experimental data are strictly the same
(multiplied by 4, in the following),
in order for the comparison to be fair.
In Figure~\ref{f:fig05}, these simulation results are plotted together with the
current-concentration curve of the present model with $N=4$.
Most of the parameters are the same as in the previous figures.
The sensitivity factor $\delta$ is adjusted to 10, and the membrane potential is fixed at $V=0.12$, in accordance with the membrane potential used in the simulations by Gordon.
Five different values of $b$ were used, namely, 0 (thin solid line), 1 (thin dashed line), 
2 (dotted line), 3 (dashed line) and 4 (solid line) respectively.
The parameters were not further optimized for fitting of the data.
However, it seems that the change of $b$ from 0 to 3 quite nicely reflects the result of
adding negatively-charged residues at critical positions in the pore.

Compared with the experiments\cite{jgp76:425,bpj101:2671},
it was found that the simulated current-concentration of the mutant channel
reproduced the characters of the experimental data well.
The ion current through the mutant channel saturated much faster 
than that of the wild type does, and at a much higher level.
However, the half-saturation concentration of the simulated mutant channel
is ``slightly higher than the the experimentally determined values for other potassium channels''
and is ``somewhat lower than that determined from KcsA using BD simulations''.
Figure~\ref{f:fig05} shows that if the initial structure is even more stabilized by
positive ions, the current level will be even higher, and the half-saturation concentration
will be lower.
In the words of the present model, those BD simulations did include important effects that
resulted in the higher ion current, but it is still not enough.
According to the present model, there may be a few possibilities.
First, they used single-ion static PMF to reflect the difference between the wild-type and
mutant channels.
As discussed earlier, this picture may not be the most appropriate for long channels like Kir.
Second, in their picture, the estimation of the energy difference between the mutant and
the wild type was, although standard, not the most deliberate for studying the electrostatic
interaction in biological systems.
It was found that the dielectric environment in the biological system tends to make the
attraction force between opposite charges weaker and the repulsion force between 
like charges stronger\cite{pre81:031925}.
Therefore, if the empty channels are considered as the initial states, and the energy is taken
as the reference point, the free energy difference between the stabilized structures of the mutant 
and wild-type channels should be even larger than predicted by standard theory.
Since the attraction force is generally weaker, it may also imply that even more positive
ions than predicted by conventional theory would be in the stable structure.
Both possibilities suggest that the $b$ and $\mu$ parameters may be higher in reality
than in the model system of Gordon.
To limit the length of discussions, the effect of $\mu$ is not systematically presented.
Just by considering the effect of increasing the value of $b$, say from 3 to 4,
does further lower the half-saturation concentration of the model.

Qualitative discussions about the correspondence between the present theory and
either other simulations or experiments were all consistent.
Nevertheless, it would be more insightful if the structural stabilities of ion-associated 
and empty ion channels can be compared quantitatively.
This is the major assertion of this work.


\section{Conclusions}\label{s:conclusion}

In this work, a model system in which positively cooperative transport processes are
coupled to negatively cooperative binding processes was proposed,
and its behavior was discussed.
Two major assertions were made, which are considered as the crucial features
that have to be appended into all existing theoretical models.
First, many channels can only be well stabilized to their native conducting structures by including
a few permeating ions into them.
Thermodynamically, this is equivalent to saying that the binding affinities of the first few
permeating ions are much higher.
It is energetically much less favored for more ions to enter the channel.
But if they happen to do so, they are likely to induce efficient ion transport through the
knock-on mechanism.
Second, since the knock-on mechanism is the most reasonable transport mechanism in
long channels, the reaction path used for studying the potential of mean force of
the ion transport process is a hyper-curve on a multi-dimensional surface.
This was already the way in which many of the simulation scientists analyzed their systems.
However, with a chemical kinetics model, the essential meanings of the complicated
potential hypersurface, the reaction path, and the potential of mean force,
can be more easily encapsulated, investigated, presented and interpreted.

The ion-blocking and unblocking process, despite being different from the
transport process in several important aspects, can be used to qualitatively discuss
the increase and decrease of ion transport rate under the change of 
the binding and structural stability characters of the system.
Moreover, the binding reactions can be assumed to operate in steady state satisfying the
pre-equilibrium approximation.
Through the careful considerations and designs, the number of arbitrary parameters can be 
kept as few as possible,
while the major features were maintained.

By providing the numerical examples of the present model and showing them
side by side with the results of some experiments and other simulations,
it was demonstrated that the considerations of this model pointed out the direction to
improve other theories and simulations in order to diminish the gap between models
and reality.
The I-V relation of model systems in molecular dynamics simulations,
and the current-concentration curve of model systems in Brownian dynamics simulations, 
were generally lower than that observed experimentally.
The present model seems to have identified the underlying problems of the simulations.


\begin{acknowledgments}
This work was financially supported by the National Science Council of Taiwan, project NSC 99-2113-M-001-021.
\end{acknowledgments}

\bibliography{cooptran}

\end{document}